\documentclass[useAMS,referee,usenatbib]{biom}
\usepackage{amsmath,amsfonts,bm,bbm,color,float,graphics,graphicx,multicol,multirow,url}
\usepackage[figuresright]{rotating}

\def\bSig\mathbf{\Sigma}

\newcommand{\ind}{\mathbbm{1}}
\newcommand{\Tref}{\mathcal{T}_{\text{ref}}}

\title[Causal Standardization]{Comparing HIV Vaccine Immunogenicity across Trials with Different Populations and Study Designs}

\author{Yutong Jin$^{1,*}$, 
Alex Luedtke$^{2}$, 
Zoe Moodie$^{3}$, 
Holly Janes$^{3}$,
David Benkeser$^{1}$\email{yutong.jin@emory.edu} \\
$^{1}$Department of Biostatistics and Bioinformatics, Emory University, Atlanta, GA, USA \\
$^{2}$Department of Statistics, University of Washington, Seattle, WA, USA \\
$^{3}$ Vaccine and Infectious Disease Division, Fred Hutchinson Cancer Research Center, Seattle, WA, USA}

\begin{document}


\date{{\it Received October} 2007. {\it Revised February} 2008.  {\it
Accepted March} 2008.}



\pagerange{\pageref{firstpage}--\pageref{lastpage}} 
\volume{64}
\pubyear{2008}
\artmonth{December}


\doi{10.1111/j.1541-0420.2005.00454.x}


\label{firstpage}


\begin{abstract}
Safe and effective preventive vaccines have the potential to help stem the HIV epidemic. The efficacy of such vaccines is typically measured in randomized, double-blind phase IIb/III trials and described as a reduction in newly acquired HIV infections. However, such trials are often expensive, time-consuming, and/or logistically challenging.
These challenges lead to a great interest in immune responses induced by vaccination, and in identifying which immune responses predict vaccine efficacy. These responses are termed vaccine \emph{correlates of protection}. Studies of vaccine-induced immunogenicity vary in size and design, ranging from small, early phase trials, to case-control studies nested in a broader late-phase randomized trial. Moreover, trials can be conducted in geographically diverse study populations across the world. Such diversity presents a challenge for objectively comparing vaccine-induced immunogenicity. To address these practical challenges, we propose a framework that is capable of identifying appropriate causal estimands and estimators, which can be used to provide standardized comparisons of vaccine-induced immunogenicity across trials. We evaluate the performance of the proposed estimands via extensive simulation studies. Our estimators are well-behaved and enjoy robustness properties. The proposed technique is applied to compare vaccine immunogenicity using data from three recent HIV vaccine trials.
\end{abstract}

%



\maketitle

\section{Introduction}
\label{sec:introduction}
Outbreaks of infectious diseases remain a major concern. In combating such diseases, the development of safe and effective preventive vaccines is crucial. 
Vaccines are often designed to generate immune responses that protect individuals against infection and/or disease caused by pathogens such as viruses, bacteria or parasites. The efficacy of candidate vaccines is typically measured in randomized, double-blind, placebo-controlled clinical trials. Vaccine efficacy (VE) is typically quantified as one minus a relative risk, comparing risk of infection or disease under vaccination to risk under a placebo or control vaccine.
However, estimation of VE against a clinical endpoint can be time-consuming, costly, and difficult to assess in randomized trials. For rare endpoints, it can take thousands of participants and years to complete such a trial. For emerging pathogens such as Chikungunya, Lassa fever, and Nipah virus, it is logistically challenging to implement randomized trials due to unpredictable and short-lived outbreaks. Moreover, randomized trials are unlikely to generate all evidence needed to guide policies around vaccines, such as whether vaccines should be updated with the emergence of new strains of a pathogen in the population.

While randomized trial-generated evidence is the gold standard for demonstrating efficacy of a vaccine, there is intense interest in identifying immune responses that are predictive of VE. Such responses, termed vaccine \emph{correlates of protection} (CoP), may serve as surrogate endpoints in lieu of a formal evaluation of clinical VE, thereby potentially opening accelerated pathways for new vaccine products to be brought to market and/or updates to the strain included in existing vaccines.
Therefore, in many contexts it is often of interest to study immunological endpoints and compare vaccine immunogenicity across different vaccines. 
The most common statistical approach to quantifying differences in immune responses across various vaccines is to use a t-test or Wilcoxon Signed-Rank test. Sometimes these simple procedures must be extended to account for the sampling design, for example, case-control studies nested in a broader randomized trial \citep{banzhoff2003new,chung2014polyfunctional,furuya2021comparison}. 

We consider the problem of extending these methods to enable more objective comparisons of vaccines when immunogenicity is evaluated in different studies across diverse geographic sites, using varying study designs. When vaccines are evaluated at different study sites, there may be important differences in clinical and/or demographic characteristics across the various trial populations. If these characteristics also impact immune responses, then simple approaches may yield biased inference regarding differences in vaccine immunogenicity. Moreover, deriving proper standard errors for estimators and tests can be challenging when comparing data generated under different designs. We make use of efficient estimators based on influence functions to tackle both of these challenges.

Our motivation for studying this problem arises from the field of HIV vaccines. Over the past decade, there have been many small and several large trials of preventive HIV vaccines. These trials have been conducted across South East Asia, sub-Saharan Africa, and in the Americas, each with their own specific set of enrollment criteria. Much of the recent work in HIV vaccine development has been motivated by the results of the RV144 trial, which demonstrated modest but significant vaccine efficacy against HIV-1 infection \citep{rerks2009vaccination}. A key consideration for HIV vaccine development is the selection of vaccine immunogens. Immunogens are molecules that are capable of eliciting a host immune response. An immunogen in the vaccine studied in the RV144 trial was designed to protect against clades circulating locally in Thailand. The RV144 trial showed modest preventive vaccine efficacy, with an estimated 31.2\% reduction in the cumulative incidence of HIV-1 infection over 42 months. These results were encouraging, led to a large correlates analysis,  and prompted several smaller follow-up trials including the HVTN 097 trial, which was designed to evaluate immunogenicity of the same vaccine regimen in South Africa \citep{gray2019immune}. The HVTN 097 results indicated that response rates and magnitudes of putatively protective immune responses in South Africa were similar or better than those observed in RV144 in Thailand, providing support for continuing with this vaccine approach for research in sub-Saharan Africa. A subsequent study, HVTN 100, evaluated a revised version of the RV144 vaccine but updated with immunogens based on HIV-1 subtype C prevalent in South Africa and a different adjuvant (MF59 in place of alum). HVTN 100 determined that the South African-adapted vaccine successfully met the pre-specified immunological criteria for advancement to efficacy testing \citep{bekker2018subtype}. Based on these results, a larger phase IIb/III randomized, double-blind, controlled efficacy trial, HVTN 702, was conducted to evaluate the vaccine efficacy of the subtype C-adapted vaccine against  HIV-1 acquisition in South Africa. Unfortunately, this trial was halted after pre-specified non-efficacy criteria were met at an interim analysis \citep{gray2021vaccine}. To help identify potential explanations for the lack of efficacy in HVTN 702 and to inform future directions for the HIV vaccine field, it is important to examine possible differences in immunogenicity between the vaccines used in HVTN 097, HVTN 100, and HVTN 702 in South Africa. \cite{moodie2022analysis} studied immune correlates of risk and concluded that the CD4+ T cell response rate in HVTN 097 to 92TH023 (74\%) was similar to that in HVTN 702 against antigen ZM96 (63\%), but that the IgG response rate to IgG 1086.C V1V2 in RV144 was significantly higher than that in HVTN 702 (100\% vs. 67\%). We developed the framework described herein to support these comparisons across different trial populations. 

The need for a comparison of vaccine-induced immune responses across vaccines that are evaluated in trials is not unique to HIV vaccines. Indeed, this is a common and important problem in many domains of vaccine research. For example, recent studies that separately evaluated the Moderna mRNA-1273 preventive COVID-19 vaccine and the Pfizer/bioNtech BNT162b2 vaccine had different sampling designs and different covariate distributions for enrolled participants \citep{baden2020efficacy, polack2020safety}. Similarly, recent studies of dengue vaccines have been conducted in diverse populations across Southeast Asia and South America. Past dengue exposure may be a key modifier of the immunogenicity and efficacy of the vaccines and circulating serotypes of dengue virus may differ across geography \citep{rabaa2017genetic, sridhar2018effect}. 

The above scientific context highlights a clear need for understanding the causal relationship between a vaccine regimen and its immunogenicity in a particular population. Such information can guide the design of new vaccines, as well as the prioritization of current vaccines for further research. In this paper, we develop a framework that identifies appropriate causal estimands and estimators that can be used to provide standardized comparisons of vaccine immunogenicity. We propose estimators of these causal estimands and establish theory that dictates the large sample behavior of the estimators. Our estimators account for two practical difficulties that arise in vaccine trials. First, we propose methodology that accounts for different sampling designs that may be used to measure immune responses across trials. For example, HVTN 100 and HVTN 097 measured immune responses in all participants; however, in HVTN 702 immune responses were measured using case-control sampling. Second, our methodology allows for pooling of trial data to gain efficiency when the same vaccine is evaluated in multiple trials. For example, an identical vaccine was evaluated in HVTN 100 and HVTN 702 and therefore, we may wish to pool data from these trials when evaluating immunogenicity. We clarify the formal causal assumptions and semiparametric efficiency theory that allows such pooling. Our work relates to other recent work on transportability of causal effects \citep{stuart2015assessing,bareinboim2016causal,li2023efficient}, focusing on the specific challenges of these approaches in the context of vaccine immunogenicity studies.

\section{Materials and Method}
\label{sec:methods}

\subsection{Notation and Data Structure}
\label{subsec: notation}
Vaccine trials can be generally categorized as being early or late phase studies. Early phase studies are often designed specifically to evaluate immunogenicity of one or several vaccine candidates and/or candidate doses of vaccine. These trials typically have smaller sample sizes, generally less than several hundred participants and do not assess VE on a clinical endpoint of interest. They may include one or several doses of a single vaccine, or one or several variations of a vaccine (e.g., vaccines with different adjuvants). We use the variable $T \in \mathcal{T}=\{1, 2, \dots, N_T\}$ to denote an arbitrary numeric label applied to the various trials considered in a particular application. Data in each of these trials contains a possibly categorical variable indicating which of the vaccine formulations/doses a participant receives denoted by the label $A \in \mathcal{A}=\{0, 1, 2, \dots, N_A\}$. A given vaccine $a \in \mathcal{A}$ could be evaluated in multiple trials; however, in our notation we use only a single, unique label for each vaccine and we denote by $\mathcal{T}_a \subseteq \mathcal{T}$ the trials in which the immunogenicity of vaccine $a$ was evaluated. The observed data also include measurements of one or several immune responses of interest $S$. In practice $S$ may be a vector, but we focus here only on scalar-valued $S$, as we can separately apply our methods to each immune response of interest. As a concrete example, we may consider HVTN 702, a phase IIb/III trial where participants were randomized to receive either an active vaccine or a placebo, and the immune responses of interest included various CD4+ T cell responses and IgG binding antibody responses.

Each trial's data will also generally include other participant-level information collected prior to vaccine assignment. The specific baseline characteristics measured may vary across trials, and we introduce $\bm{W}(t)$ to denote covariates measured in trial $t = \{1,2, ..., \tau\}$. We use $\bm{W} = \bm{W}(1) \cup \bm{W}(2) \cup ... \cup \bm{W}(\tau)$ to denote the superset of covariates consisting of all covariates collected in at least one of the trials considered. In our HIV vaccine example, participants in HVTN 097, 100 and 702 had their age, gender, body mass index (BMI), region of enrollment and educational level recorded. Thus, in this example $\bm{W}$ would include five demographic variables that were available in three trials.

In addition to vaccine, immune response, and covariates, some trials will also have data available on clinical endpoints of interest. This will almost always be the case for larger phase IIb/III trials that are designed explicitly to evaluate VE. For example, in HVTN 702, the primary outcome was time to first detection of HIV-1 acquisition or censoring and this information is recorded for all participants in the trial.
Thus, we can assume that for some trials, the observed data will also include a clinical outcome of interest, which we denote by $Y$. The outcome could be binary (e.g., indicator of disease by a fixed time-point) or it may be a time-to-event endpoint (e.g., time since vaccination until first occurrence of clinical disease). Our methods apply readily to both situations; however, for simplicity we hence assume $Y$ is binary. In early phase trials, $Y$ may be missing or right-censored for most or all individuals. This missingness pattern has no adverse impact on our developments, since we are primarily interested in comparing $S$ across vaccines and $Y$ typically occurs after $S$. If $Y$ is subject to missingness it will be entirely appropriate for our developments to consider this variable as a three-level categorical variable with levels 0, 1, and missing.

An interesting aspect of the design of many vaccine trials is that $S$ may not be measured on every participant. Therefore, we introduce two versions of the data structure that allow us to differentiate between settings where $S$ is measured on every participant in every trial and settings where $S$ is only measured on a subset of participants in at least some of the trials considered. We refer to a datum collected in the former setting as a \emph{full data unit} and in the latter setting as a \emph{observed data unit}. Explicitly considering the full data unit in this setting is useful mathematically for describing requisite assumptions for identification of our causal estimands of interest. In the full data setting, for each participant in each trial, we record $X=(T,A,\bm{W},S,Y) \sim P_X$, while recalling that without loss of generality, $Y$ will generally be coded as right-censored or missing for most or all individuals enrolled in early phase trials. In our notation, we use $P_X$ to denote the probability distribution $X$, which is assumed to follow a statistical model $\mathcal{M}_X$ that is nonparametric up to certain assumptions detailed in Section \ref{sec:identification_full} below. We use $E_X$ to denote expectation of a random variable under sampling from $P_X$.

We now turn to the \emph{observed data unit}, where $S$ may not be measured on all individuals.
Many phase IIb/III trials employ a \emph{two-stage sampling design} to efficiently determine the subset of participants in which $S$ should be measured \citep{breslow2005}. In these designs, all participants have specimens (e.g., serum) collected at a clinic visit following vaccination but immune responses are only measured for participants in the subset. For example, in HVTN 702, a case-control design was used, wherein all vaccinated participants assigned female at birth who tested positive for HIV-1 after the month 6.5 study visit but before the end of primary follow-up (month 24) were selected and measured for $S$ at months 6.5 (and 12.5 for those who acquired HIV-1 thereafter). A covariate-matched set of controls was also selected to have $S$ measured. Hence, out of the 1168 female per-protocol participants eligible for case-control sampling in HVTN 702, $S$ was measured in 130 of these individuals \citep{moodie2022analysis}. To accommodate the potential for the presence of two-stage sampling, we introduce the \emph{observed data unit} $O=(T,A,\bm{W},\Delta, \Delta S, Y) \sim P$, which is a \emph{coarsened} version of the full data unit $X$. A typical observed data unit includes $T$, $A$, $\bm{W}$ and $Y$ (possibly subject to missingness) as above; however, the immune response $S$ is measured only in a subset of participants. The random variable $\Delta$ takes value 1 if the immune response $S$ is measured and zero otherwise. In the data unit $O$, without loss of generality we represent the observed value of $S$ by $\Delta S$, thereby arbitrarily recording a value of 0 for $S$ in individuals not selected for two-phase sampling. We note that for early phase trials generally we will have $\Delta = 1$ for all participants, while for late phase trials, $\Delta = 1$ for only a subset of participants. The statistical model $\mathcal{M}$ for $P$ is implied by the model for the distribution of the full data unit $P_X$ and the model for the sampling variable $\Delta$ given $(T,A,W,Y)$, where these sampling probabilities are generally known by design. We use $E$ to denote the expectation of random variable under sampling for $P$.

We provide an example visualization of the type of observed data that is used in our motivating example in Supplementary Table 1. In this example, our data set consists of data from three trials pooled into a single data set. Our two covariates $\bm{W}$ of interest are categorical age ($W_1$) and sex at birth ($W_2$). 
There is one late phase trial, HVTN 702 and two early phase trials HVTN 100 and HVTN 097. In the late phase trial, HIV-1 acquisition $Y$ is recorded for all participants (for simplicity, we present an idealization of the actual data where we ignore right-censoring of $Y$). However, the immune response $S$ of interest is only measured in a subset of participants in these trials, as indicated by rows where $\Delta = 1$; $S$ is missing for all rows in which $\Delta = 0$. On the other hand, in the early phase trials, $S$ is measured for everyone, while $Y$ is generally missing. At times, we will refer back to Supplementary Table 1 to make concrete our general estimation strategies.

\subsection{Causal Estimands}
\label{subsec:causal_estimand}
Traditionally, the average immune response induced by each vaccine is estimated in each trial separately. This approach targets the estimand $\mu_{a} := E_X(S \mid A=a, T=t)$. However, in some situations there may be components of $\bm{W}$ that are correlated with both trial enrollment $T$ and immune responses $S$. For example, age may vary across trials and correlate with the magnitude of vaccine-induced immune responses, rendering the comparison of two vaccine candidates $\mu_{a}-\mu_{a'}$ evaluated in different trials $t$ and $t'$ biased.

To address this concern, we propose a causal framework to provide such comparisons in an appropriate way. In particular, we can consider a counterfactual variable $S(a)$ that corresponds to the immune response that would be observed if an individual were given vaccine $a$. We assume that causal consistency holds and that there is no interference between individuals. Both assumptions are reasonable in the present context, where causal consistency stipulates that there are not ``multiple formulations'' of a single vaccine. This assumption is generally reasonable for most vaccines, where often a key goal of pre-clinical vaccine development process is developing consistent manufacturing processes to ensure comparable vaccines across lots. No interference is also likely to be plausible in the present context as the immune response of one individual is unlikely to depend on vaccines received by other individuals in the study.

In this counterfactual scenario, it is possible for all individuals who could potentially enroll in any of the trials to receive any of the $N_A$ vaccines considered. Thus, we can conceptualize a counterfactual data unit $\mathbb{X} = (T, \bm{W}, \{S(a), Y(a) : a \in \{1, \dots, N_A\}\}) \sim P_{\mathbb{X}}$, where for completeness we define $Y(a)$ as the counterfactual clinical endpoint that would be observed under vaccination with $a$, though this quantity does not play a role in our development. As above, we denote by $E_{\mathbb{X}}$ expectation of a random variable under $P_{\mathbb{X}}$.

We are ultimately interested in comparing immunogenicity, for example, by comparing the average value of $S(a)$ vs. $S(a')$ for vaccines $a, a'$ that were evaluated in different trials. However, when these vaccines are evaluated in different trials that enroll from different populations, there are several such comparisons that could be of interest. In the context of HIV vaccines, a series of trials were conducted in several countries across several years. As described in the introduction, in our motivating example, the population of HVTN 702 was of primary interest, as our goal is to compare the immunogenicity across vaccines to aid in the interpretation of the null signal in the primary vaccine efficacy analysis of the HVTN 702. Thus, we may be interested in understanding whether and how the immunogenicity of the vaccine formulation studied in the earlier HVTN 097 trial compares to the formulation studied in HVTN 702, while making this comparison \emph{in the HVTN702 trial population}. That is, we are asking a hypothetical question about the immunogenicity that \emph{would have been observed} had we evaluated the HVTN 097 vaccine alongside the HVTN 702 vaccine, \emph{in the HVTN 702 study population}. Using the labels from Supplementary Table 1, 
this estimand would be denoted $E_{\mathbb{X}}[S(2) - S(1) \mid T = \text{HVTN702}]$. We label this type of causal estimand a \emph{standardized} comparison of immunogenicity. 

While our motivating example focuses on a setting where a single trial's population is of interest, more generally we could consider standardized comparisons of the form $E_{\mathbb{X}}[S(a) - S(a') \mid T \in \Tref]$, where $\Tref \subseteq \mathcal{T}$ may include multiple trials. We refer to $\Tref$ as the \emph{referent trial(s)} to which we are standardizing our comparison. The choice of referent trial should be dictated by the scientific context. While we generally expect that $\Tref$ will consist of a single trial, in some situations we may wish to include multiple trials in our referent. For example, if vaccines $a$ and $a'$ are evaluated in trials that enroll from very similar or identical populations, then we may wish for $\Tref = \mathcal{T}_a \cup \mathcal{T}_{a'}$. A trivial situation where this might occur is when vaccines $a$ and $a'$ are evaluated in the same trial and we are interested in inference on their immunogenicity in that trial's study population. However, we may also have settings where vaccines are evaluated in different trials, but the distribution of common baseline covariates are largely similar among trials $\mathcal{T}_a$ and $\mathcal{T}_{a'}$. This could happen when vaccines are evaluated at the same study sites using trials with similar enrollment criteria. 
In the absence of effect heterogeneity by covariates, inference on this quantity may have greater precision than inference based on an estimand standardized to either $\mathcal{T}_a$ or $\mathcal{T}_{a'}$ alone.

\emph{Remark}: Another potential setting is one where there exists a common referent population that is not sampled directly from any of the observed trials. For example, we may wish to compare immunogenicity of two vaccines in an age- and sex-standardized way against a known referent population distribution. We provide theory for this estimation and inference pertaining to this estimand in the Supplementary C. 

\subsection{Identification of standardized immunogenicity using full data} \label{sec:identification_full}

To identify the standardized immunogenicity comparison described in our motivating example, we require certain \emph{causal} assumptions regarding the distribution of $\mathbb{X}$, in addition to assumptions pertaining to the sampling design of $S$ as encoded in the distribution of $O$. Key to both sets of assumptions is the consideration of which baseline covariates are available across the various trials.

We introduce the general notation $\bm{W}_{\cap}(\mathcal{T}_0)$ to denote covariates that are available across \emph{all} of a given set of trials $\mathcal{T}_0 \subseteq \mathcal{T}$. Thus,  $\bm{W}_{\cap}(\Tref) \subseteq \bm{W}$ refers to the covariates common to all referent trial(s) and $\bm{W}_{\cap}(\mathcal{T}_a)$ denotes covariates common to all trials where vaccine $a$ is evaluated.
We denote by $\bm{W}_{\cap}(\mathcal{\Tref}) \cap \bm{W}_{\cap}(\mathcal{T}_a)$ the set of covariates that are available in all referent trials and all trials where vaccine $a$ is evaluated. This set of covariates is particularly important for identification. As we presently show, we must be able to identify a subset of these covariates that is sufficient to control for differences in counterfactual immunogenicity between the individuals receiving vaccine $a$ and individuals in the referent population. We make the simplifying assumption that such covariates \emph{must} be available in all of the trials where the immunogenicity of vaccine $a$ is actually measured, so that we can identify the vaccine's expected immunogenicity conditional on this set of covariates. Moreover, we also need the \emph{same set of covariates} to be available in the referent trial(s) so that the covariate-conditional immunogenicity can be properly standardized to the referent trial. 
We further discuss the identification under a weaker assumption that covariates are only available in at least one trial where vaccine $a$ is evaluated (Supplementary F).

Formally, identification of $\psi_{\mathbb{X}}(a) = E_{\mathbb{X}}[S(a) \mid T \in \Tref]$ for an arbitrary vaccine $a$ using the full data requires the following assumptions. 
\begin{enumerate}
    \item[(A1)] \emph{Ignorability of trial enrollment and vaccine assignment conditional on common covariates.} There exists a set of common baseline covariates $\bm{W}_S \subseteq \bm{W}_{\cap}(\mathcal{\Tref}) \cap \bm{W}_{\cap}(\mathcal{T}_a)$ such that: (A1.1) $S(a) \perp A \mid T \in \Tref, \bm{W}_S$ and (A1.2) $S \perp T \mid A, \bm{W}_S$.
    \item[(A2)] \emph{Positivity of vaccine assignment.} $P_X\{ P_X(A=a \mid \bm{W}_S) > 0 \mid T \in \Tref \}=1$
\end{enumerate}
\vspace*{-\baselineskip}
Assumption (A1.1) stipulates that we must be able to identify a set of covariates that are measured in both the referent trial and the trial(s) where vaccine $a$ is evaluated such that conditional on this set of covariates, the vaccine which a referent-trial participant is observed to receive provides no additional information about their potential outcome $S(a)$. This condition will generally hold by design if vaccines are randomly assigned, but may require additional scrutiny in observational studies. Assumption (A1.2) stipulates that conditional on vaccine assignment $A$ and $\bm{W}_S$, the particular trial provides no additional information about the immune response outcome $S$. Generally, we can think about two sub-assumptions that are needed to satisfy this assumption. First, there can not be a direct effect of trial on vaccine immunogenicity. This assumption would be violated if, for example, a trial had inappropriate cold storage procedures thereby causing weakened immunogenicity of the vaccine. Second, we require that $\bm{W}_S$ includes all characteristics that are related to both vaccine immunogenicity and that may differ across trial populations. For example, consider a scenario where certain compositions of the gut microbiome have a positive impact on vaccine immunogenicity and microbiome data are not available as part of $\bm{W}_S$. If microbiome composition differs across trials in $\mathcal{T}_a$, then assumption (A1.2) would be violated. Graphical approaches may be useful for scrutinizing this assumption in each specific scientific context.
We remark that our notation $\bm{W}_S$ indicates that the choice of covariates may differ depending on the immune response that is being studied, as different responses may have different biological drivers. The choice of covariates $\bm{W}_S$ may also differ depending on the particular vaccine $a$ and the particular choice of referent trial $\Tref$. However, for simplicity we have elected to suppress this dependency in our notation.
Assumption (A2) stipulates that there is a positive probability of receiving vaccine $a$ for all values of $\bm{W}_S$ that are observable in $\Tref$. This assumption would be violated if, for example, there were certain combinations of covariates that are observable in the referent trials $\Tref$, but not in any of the trials in which vaccine $a$ was studied. This condition could be scrutinized empirically using standard methods for evaluating propensity score overlap, for example by evaluating an estimate of $P_X(A = a \mid \bm{W}_S)$ using observations in $\Tref$ \citep{austin2015moving}.\vspace*{-1em}


\begin{theorem}
\label{thm:full_data_id}
If (A1) and (A2) hold, then $\psi_\mathbb{X}(a) = E_X\left[ E_X(S \mid A=a, \bm{W}_S) \mid T \in \Tref \right]$. 
\end{theorem}
A detailed proof can be found in Supplementary B. We hence use $\psi_X(a) = E_X[ E_X(S \mid A=a, \bm{W}_S) \mid T \in \Tref ]$ to refer to the identifying estimand as distinct from the causal estimand $\psi_{\mathbb{X}}(a)$. The implication of Theorem \ref{thm:full_data_id} is that if (A1) and (A2) hold then $\psi_{\mathbb{X}}(a) = \psi_X(a)$ and a causal standardized immunogenicity comparison is possible using data sampled from $P_X$. However, even in the ideal context where $\mathcal{T}_a$ consists of only randomized trials, assumption (A1.2) may yet be considered unreasonable. For example, the various trials in $\mathcal{T}_a$ may collect different sets of key covariates, rendering this assumption difficult to satisfy based on the set of covariates common to $\Tref$ and $\mathcal{T}_a$. In this case, $\psi_X(a)$ does not have a causal interpretation. Nevertheless, we argue that a comparison of $\psi_X(a)$ and $\psi_X(a')$ still retains a useful non-causal interpretation as a covariate-adjusted comparison of vaccines $a$ and $a'$, standardizing the set of available common covariates to their distribution in the referent trial(s). So long as $\bm{W}_S$ contains at least some covariates that are prognostic of immune response and whose distributions differ across trials, we argue that a comparison of $\psi_X(a)$ and $\psi_X(a')$ may still be preferred over a na{\"i}ve estimand that compares $\mu_{a}$ and $\mu_{a'}$ directly.

\subsection{Identification of standardized immunogenicity using observed data}
We now describe how $\psi_X(a)$ can be identified in the coarsened data setting, where we are sampling data from $P$ rather than $P_X$. 
In a particular trial $t$, sampling probabilities for $S$ could be determined based on $A$ (e.g., we may over-sample vaccine recipients and under-sample placebo recipients), $Y$ (e.g., it is common to sample all cases and only a subset of the remaining trial participants), and/or a subset of available covariates $\bm{W}(t)$ (e.g., we may over-sample particular populations to ensure appropriate representation in the observed data). We denote by $\bm{W}_\Delta(t) \subseteq \bm{W}(t)$ the set of covariates, if any, that are used to determine sampling probabilities in trial $t$. 
Here, to simplify the exposition, we make the simplifying assumption that all trials in 
$\mathcal{T}_{a}$ 
use two-stage sampling and that the covariates used for sampling, $\bm{W}_\Delta(t)$ are the same for all such trials. We refer to this set of covariates as $\bm{W}_{\Delta}$, suppressing the dependence on $a$ for simplicity. 
In future work, we will demonstrate how this assumption may be relaxed to allow different sampling designs across 
$\mathcal{T}_{a}$; we expect this generalization to be straightforward.

To identify $\psi_X(a)$ using the observed data we require the following assumptions.
\begin{enumerate}
    \item[(A3)] \emph{Missing at random}: $S \perp \Delta \mid T, A, \bm{W}_{\Delta}, Y$
    \item[(A4)] \emph{Positivity of sampling}: $\forall \ t \in \mathcal{T}_a$, $P\{ P(\Delta = 1 \mid t, a, \bm{W}_{\Delta}, Y) > 0 \mid A=a, T=t \}=1$.
\end{enumerate}

Assumption (A3) stipulates that given $(T, A, \bm{W}_{\Delta}, Y)$ the probability of having immune responses measured cannot depend on the underlying immune response itself. Sampling probabilities are generally selected a-priori by design in late phase vaccine trials, so we expect this assumption will typically be satisfied. If instead, trials are designed such that immune responses are measured subject to some form of convenience sampling (e.g., participants can self-select into an immunogenicity sub-study), then this assumption would require further scrutiny. Assumption (A4) stipulates that there is a positive probability of sampling immune responses for measurement for each available covariate profile in each trial where vaccine $a$ is administered. Again, this assumption can generally be ensured by design.
We define $\bm{W}_{\Delta,S} = \bm{W}_S \cup \bm{W}_{\Delta}$ to be the union of covariates required to satisfy (A1) and (A3). We have the following identification result for $\psi_X(a)$.
\begin{theorem} \label{thm:obs_data_id}
If Assumptions (A3)-(A4) hold then 
$\psi_X(a) = E\{ E[ E(S \mid \Delta = 1, A=a, T \in \mathcal{T}_{a}, Y, \bm{W}_{\Delta,S}) \mid A = a, T \in \mathcal{T}_{a}, \bm{W}_S ] \mid T \in \Tref\}$.
\end{theorem}

\subsection{Towards estimation: efficiency theory for identifying estimands}

In this section, we provide the efficient influence function (EIF) of $\psi_X(a)$ and $\psi(a)$ in models that assume (A1)-(A4). We recall that an estimator's \emph{influence function} is a function of the data unit that has mean zero and finite variance. In particular, an estimator $\psi_n(a)$ of $\psi(a)$ is said to have influence function $D$ if $\psi_n(a) = \psi(a) + n^{-1} \sum_{i=1}^n D(O_i) + o_P(n^{-1/2})$. Influence functions are particularly useful for so-called \emph{regular} estimators, as they can also be used to characterize the efficiency bound of all such estimators of a given parameter. The influence function of the regular estimator with the smallest asymptotic variance is called the efficient influence function. Influence functions are often indexed by so-called \emph{nuisance parameters}, parameters of the data generating distribution that are not directly of interest, but are useful for constructing and studying the large sample behavior of estimators of the estimand of interest. Thus, influence functions can provide a means to generate final estimates with desirable large sample behavior \citep{rose2011targeted}. 

We introduce some additional notation to represent the nuisance parameters indexing our efficient influence function. We define $\bar{Q}_X(\bm{W}_{i,S}) = E_X(S \mid A = a, \bm{W}_S = \bm{W}_{i,S})$ as the conditional mean immune response, $g_A(a \mid \bm{W}_{i,S}) = P_X(A = a \mid \bm{W}_S = \bm{W}_{i,S})$ as the conditional probability of vaccine $a$ given covariates $\bm{W}_{i,S}$, $g_T(\mathcal{T}_0 \mid \bm{W}_{i,S}) = P_X(T \in \mathcal{T}_0 \mid \bm{W}_{S} = \bm{W}_{i,S})$ as the conditional probability of enrollment in one of the trials in a given set $\mathcal{T}_0 \subseteq \mathcal{T}$ given covariates $\bm{W}_{i,S}$, 
and $g_T(\Tref) = P_X(T \in \Tref)$ as the marginal probability of enrollment in one of the trials in $\Tref$.
\vspace*{-\baselineskip}
\begin{theorem}
\label{theorem3}
    The EIF for $\psi_X(a)$ in a model for $P_X$ that only assumes (A1)-(A2) is \begin{align*}
    D_X(X_i) &= \frac{\ind_a(A_i)}{g_A(a \mid \bm{W}_{i,S})} \frac{g_T(\Tref \mid \bm{W}_{i,S})}{g_T(\Tref)} \left\{ S_i - \bar{Q}_X(\bm{W}_{i,S}) \right\} + \frac{\ind_{\Tref}(T_i)}{g_T(\Tref)} \left\{ \bar{Q}_X(\bm{W}_{i,S}) - \psi_X(a) \right\} \ .
    \end{align*}
\end{theorem}

We can also define the EIF for $\psi(a)$ (see Supplementary G), which is indexed by the following additional nuisance parameters: \begin{align*}
\bar{Q}_2(Y_i, \bm{W}_{i,\Delta,S}) &= E(S \mid \Delta = 1, A = a, T \in \mathcal{T}_{a}, Y = Y_i,  \bm{W}_{\Delta,S} = \bm{W}_{i, \Delta,S}) \ , \\
\bar{Q}_1(\bm{W}_{i,S}) &= E[ \bar{Q}_2(Y, \bm{W}_{\Delta, S}) \mid A = a, T \in \mathcal{T}_{a}, \bm{W}_S = \bm{W}_{i,S}] \ , \\
g_\Delta(1 \mid T_i, A_i, Y_i, \bm{W}_{i, \Delta,S}) &= P(\Delta = 1 \mid T = T_i, A = A_i, Y = Y_i, \bm{W}_{\Delta, S} = \bm{W}_{i, \Delta, S}) \ . 
\end{align*}
\vspace{-3em}
\begin{theorem}
    The EIF for $\psi(a)$ in a model for $P$ that assumes (A1)-(A4) is \begin{align*}
    D(O_i) &= \frac{\ind_1(\Delta_i)}{g_\Delta(1 \mid T_i, A_i, Y_i, \bm{W}_{i, \Delta,S})}\frac{\ind_a(A_i)}{g_A(a \mid \bm{W}_{i,S})} \frac{g_T(\Tref \mid \bm{W}_{i,S})}{g_T(\Tref)} \{ S_i - \bar{Q}_2(Y_i, \bm{W}_{i,\Delta,S}) \} \\
    &+ \frac{\ind_a(A_i)}{g_A(a \mid \bm{W}_{i,S}) } \frac{g_T(\Tref \mid \bm{W}_{i,S})}{g_T(\Tref)} \{ \bar{Q}_2(Y_i, \bm{W}_{i,\Delta,S}) - \bar{Q}_1(\bm{W}_{i,S}) \} + \frac{\ind_{\Tref}(T_i)}{g_T(\Tref)} \{ \bar{Q}_1(\bm{W}_{i,S}) - \psi(a) \} \ .
    \end{align*}
\end{theorem}

\subsection{Targeted minimum loss estimation}
The form of the efficient influence function suggests a natural targeted minimum loss-based estimation (TMLE) approach involving sequential regression \citep{van2006targeted}. TMLE in general consists of two major steps that are sometimes implemented iteratively. In the first step, estimators of \emph{nuisance parameters} indexing the efficient influence function are obtained. TMLE is agnostic as to how such parameters are estimated, though regression stacking or super learning is commonly used towards this end in practice \citep{van2007super}. The second step of TMLE improves by using empirical risk minimization in a low-dimensional parametric model to simultaneously (i) improve the fit of initial nuisance parameter estimates and (ii) ensure that, at the end of the TMLE procedure, the so-called \emph{efficient influence function estimating equation} is solved. 

A TMLE for $\psi(a)$ may be implemented in the following specific steps. Further information pertaining to hypothesis testing is available in Supplementary Section H.

\begin{enumerate}

    \item \textbf{Estimate the probability of enrollment in referent trial(s) given covariates.} To estimate $g_T(\Tref \mid \bm{W}_S)$, we can use data from all observed trials to fit a regression with the outcome $\ind_{\Tref}(T)$ and predictors $\bm{W}_S$. This regression could be estimated using any binary regression approach. Denote the estimate by $g_{n,T}$ and define the estimated marginal probability of enrollment in $\Tref$ as $g_{n,T}(\Tref) = n^{-1} \sum_{i=1}^n \ind_{\Tref}(T_i)$.
    
    \item \textbf{Estimate the pooled probability of receiving vaccine $a$ given covariates.}  To estimate $g_A(a \mid \bm{W}_S)$, we can use all the data to fit a regression with outcome $\ind_{a}(A)$ and predictors $\bm{W}_S$. This regression could be estimated using approach that is appropriate for binary outcome regression. Denote the estimate by $g_{n,A}$. 

    \item \textbf{Compute sampling probabilities for each individual.} Next, we need to evaluate sampling probabilities $g_{\Delta}(1 \mid T_i, A_i, Y_i, \bm{W}_{i,\Delta,S})$ for each individual who received vaccine $a$ and who were enrolled in one of the trials included in $\mathcal{T}_{a}$. These probabilities are generally known by design; if unknown, then they could be estimated separately for each trial in $\mathcal{T}_{a}$ using regression of the binary outcome $\Delta$ on predictors $A, Y, \bm{W}_{\Delta,S}$. Denote by $g_{n,\Delta}$ the estimates (or true values) of the conditional sampling probabilities.
    
    \item \label{step:first_or} \textbf{Estimate vaccine-specific conditional mean immunogenicity given sampling and other covariates.}
    To obtain an estimate of $\bar{Q}_2$, we can use data from individuals who received vaccine $a$ across all trials in $\mathcal{T}_{a}$ to fit a regression with the outcome $S$ and predictors $Y, \bm{W}_{\Delta, S}$. As above, any suitable regression technique can be used and we denote by $\bar{Q}_{n,2}$ the estimate of the conditional mean immunogenicity.
    
    \item \label{step:first_fluctuate} \textbf{Target the vaccine-specific conditional mean immunogenicity given sampling and other covariates.}
    For simplicity, suppose $S \in (0,1)$. If this assumption does not hold, then $S$ can be re-scaled to fall in this interval and the same approach can be adopted \citep{gruber2010targeted}. Using all individuals with measured immune response $\Delta = 1$ that received vaccine $a$ in trials $\mathcal{T}_{a}$, fit a logistic regression with outcome $S$, an offset equal to $\mbox{logit}[\bar{Q}_{n,2}(Y, \bm{W}_{\Delta, S})]$, and a single covariate, defined as \[
    H_{n, 2}(T, A, Y, \bm{W}_{\Delta, S}) = \frac{g_{n,T}(\Tref \mid \bm{W}_{S})}{g_{n,\Delta}(1 \mid T, A, Y, \bm{W}_{\Delta,S}) g_{n,A}(a \mid \bm{W}_{S}) g_{n,T}(\Tref) } \ .
    \]\vspace*{-0.5\baselineskip}
    \item[] Note that this regression model for $\bar{Q}_2$ has a single coefficient $\beta_2$ and the model can be expressed as $\bar{Q}_{2, \beta_2} = \mbox{expit}[\mbox{logit}(\bar{Q}_{n,2}) + \beta_2 H_{n,2}], \beta_2 \in \mathbb{R}.$ Let $\beta_{n,2}$ denote the maximum likelihood estimate of $\beta_2$ and define $\bar{Q}_{n,2}^*$ as an estimate of $\bar{Q}_{2,\beta_{2}}$.
    
    \item \label{step:second_or} \textbf{Estimate vaccine-specific conditional mean immunogenicity excluding sampling covariates.} To estimate $\bar{Q}_1$, we regress the pseudo-outcome $\bar{Q}_{n,2}^*(T, A, Y, \bm{W}_{\Delta,S})$ onto $\bm{W}_S$ using observations that received vaccine $a$. As above, any suitable regression technique can be used and we denote by $\bar{Q}_{n,1}$ the estimate of conditional mean immunogenicity, now conditioning \emph{only} on baseline covariates $\bm{W}_S$.

    \item \textbf{Target the vaccine-specific conditional mean immunogenicity excluding sampling covariates.} For simplicity, we again suppose that our initial estimates obtained in the previous step are such that $\bar{Q}_{n,1}(\bm{W}_S) \in (0,1)$ for all $\bm{W}_S$, while re-scaling can again be applied as needed. Now, using \emph{all} individuals that received vaccine $a$, fit a logistic regression with outcome $\bar{Q}_{n,2}^*(T, A, Y, \bm{W}_{\Delta,S})$, an offset equal to $\mbox{logit}[\bar{Q}_{n,1}(\bm{W}_{S})]$, and a single covariate, defined as \vspace*{-0.5\baselineskip}\[
    H_{n, 1}(\bm{W}_{S}) = \frac{g_{n,T}(\Tref \mid \bm{W}_{S})}{g_{n,A}(a \mid \bm{W}_{i,S}) g_{n,T}(\Tref) } \ .
    \] 
    Note that this regression model for $\bar{Q}_1$ has a single coefficient $\beta_1$ and the model can be expressed as $\bar{Q}_{1, \beta_1} = \mbox{expit}[\mbox{logit}(\bar{Q}_{n,1}) + \beta_1 H_{n,1}], \beta_1 \in \mathbb{R}.$ Let $\beta_{n,1}$ denote the maximum likelihood estimate of $\beta_1$ and define $\bar{Q}_{n,1}^*$ as an estimate of $\bar{Q}_{1,\beta_{1}}$.
    
    \item \textbf{Construct the final TMLE estimate}. The final estimate is 
    $$\psi_n^*(a) = \frac{1}{\sum_j \ind_{\Tref}(T_j)} \sum_{i=1}^n \ind_{\Tref}(T_i) \bar{Q}_{n,1}^*(\bm{W}_{i,S})\ .$$
\end{enumerate}


\section{Simulation Studies}
\label{sec:simulation}
We evaluated the proposed estimators via simulation studies in terms of their bias, variance, mean squared error (MSE), coverage probability of 95\% Wald-type confidence intervals and mean width of confidence intervals.

In the first simulation, we wished to compare the immunogenicity of vaccines evaluated in two different studies where the studies were imbalanced on key covariates. In this context, we considered three scenarios: (i) comparing vaccines evaluated in separate early phase trials; (ii) comparing vaccines where one is evaluated in an early phase trial and the other in a late-phase trial that used two-phase sampling for measurement of immune responses; (iii) comparing vaccines evaluated in separate late phase trials that both used two-phase sampling for measurement of immune responses. 
In each setting,  we simulated two binary covariates $W_1 \mid T \sim Bernoulli(0.65 + 0.15\ind_2(T))$ and $W_2 \mid T \sim Bernoulli(0.5 - 0.2\ind_2(T))$. We use $A = 1$ to denote the vaccine evaluated in trial $T = 1$ and $A = 2$ to denote the vaccine evaluated in trial $T = 2$. Both trials 1 and 2 were simulated to have 1:1 randomization to either their respective active vaccines or a control vaccine (arbitrarily labeled $A = 3$). The immune response was simulated as $S \mid A, \bm{W} \sim \mbox{Normal}((W_1 - W_2 + 2\ind_{\{1,2\}}(A)), 1)$ and the clinical outcome $Y$ was simulated as $Y \mid S,A,W \sim \mbox{Bernoulli}(\text{expit}(-2+\ind_{\{1,2\}}(A)+W_1/2-S/2))$. To simulate two-phase sampling, we allowed sampling probabilities to depend on vaccine $A$ and outcome $Y$. In early-phase trials $P(\Delta = 1 \mid A = a, Y = y) = 1$ for all $a, y$, consistent with the standard design of measuring immunogenicity in all participants in such trials. For late-phase trials, we generated data with two-phase sampling according to probabilities listed in Table \ref{tab:simple_sim}.

\begin{table}
\caption{\textbf{Details of generating scheme for each simulated trial set.} $n$ is the sample size and $P_{y,a}$ is the sampling probability in the sub-population $Y=y$ and $A=a$}
\label{tab:simple_sim}
\centering
\begin{tabular}{cccccccccc}
\hline
Scenario & Trial & Vaccine & $n$ & $P(W_1)$ & $P(W_2)$ & $P_{1,0}$ & $P_{1,1}$ & $P_{0,0}$ & $P_{0,1}$\\
\hline
1 & 1 & 1 & 200 & 0.65 & 0.80 & 1 & 1 & 1 & 1\\
  & 2 & 2 & 150 & 0.5 & 0.30 & 1 & 1 & 1 & 1\\ 
\hline
2 & 1 & 1 & 5000 & 0.65 & 0.80 & 0.05 & 0.1 & 0.05 & 0.1\\
  & 2 & 2 & 150 & 0.5 & 0.30 & 1 & 1 & 1 & 1\\ 
\hline
3 & 1 & 1 & 2000 & 0.65 & 0.80 & 0.05 & 0.1 & 0.05 & 0.1\\
  & 2 & 2 & 1500 & 0.5 & 0.30 & 0.05 & 0.1 & 0.05 & 0.1\\ 
\hline
\end{tabular}
\end{table}

The resultant two trial sets are stacked as a whole set and are used for two target parameter estimations via TMLE. This procedure is repeated 1000 times for each scenario and then summarized by comparing with the corresponding true values. Our estimators exhibited low bias and reasonable MSE in all proposed scenarios (Table~\ref{tab:simple_sim_result}). They also achieve nominal confidence interval coverage with reasonable width of confidence intervals.

\begin{table}
\caption{\textbf{Bias, variance, mean-squared error (MSE), coverage probability and width of 95\% CI for first simulation}. Simulation results of three scenarios are summarized for two choices of referent populations. Our methods have consistent performance with small biases, low MSE and well-defined coverage probability of $95\%$ confidence intervals. $\text{CI}_c$: CI coverage; $\text{CI}_w$: CI width.}
\label{tab:simple_sim_result}
\centering
\begin{tabular}{ccccrrrrr}
  \hline
Case & $\Tref$ & Vaccine & Truth & Bias & Variance & MSE & $\text{CI}_c$ & $\text{CI}_w$ \\
  \hline
1 & $\{1, 2\}$ & 1 & 2.0000 & 0.0014 & 0.0081 & 0.0080 & 0.9450 & 0.3513 \\ 
  1 & $\{1, 2\}$ & 2 & 2.0000 & 0.0005 & 0.0112 & 0.0109 & 0.9500 & 0.4122 \\ 
  1 & $\{1\}$ & 1 & 1.8500 & 0.0001 & 0.0069 & 0.0065 & 0.9540 & 0.3262 \\ 
  1 & $\{1\}$ & 2 & 1.8500 & -0.0004 & 0.0184 & 0.0180 & 0.9490 & 0.5267 \\ 
   \hline
2 & $\{1, 2\}$ & 1 & 1.8600 & -0.0023 & 0.0041 & 0.0043 & 0.9450 & 0.2497 \\ 
  2 & $\{1, 2\}$ & 2 & 1.8600 & 0.0047 & 0.0161 & 0.0162 & 0.9370 & 0.4925 \\ 
  2 & $\{1\}$ & 1 & 1.8500 & -0.0022 & 0.0041 & 0.0042 & 0.9430 & 0.2497 \\ 
  2 & $\{1\}$ & 2 & 1.8500 & 0.0050 & 0.0167 & 0.0168 & 0.9360 & 0.5010 \\ 
   \hline
3 & $\{1, 2\}$ & 1 & 2.0000 & -0.0008 & 0.0134 & 0.0139 & 0.9450 & 0.4493 \\ 
  3 & $\{1, 2\}$ & 2 & 2.0000 & -0.0016 & 0.0196 & 0.0214 & 0.9170 & 0.5418 \\ 
  3 & $\{1\}$ & 1 & 1.8500 & -0.0015 & 0.0108 & 0.0102 & 0.9520 & 0.4057 \\ 
  3 & $\{1\}$ & 2 & 1.8500 & -0.0016 & 0.0311 & 0.0347 & 0.9150 & 0.6789 \\ 
  \hline
\end{tabular}
\end{table}

Additional simulation studies evaluating our estimators are provided in Supplementary Sections D and E. These studies provide further evidence supporting the applicability of our method across a variety of contexts. 
The proposed method yields well-performed estimation and inference in a range of vaccine trial settings, consistently exhibiting low bias, small MSE and well-calibrated 95\% Wald-type confidence intervals for our estimators.

\section{Application to HVTN Trials}
\label{sec:data_analysis_hiv_ics}
The proposed methods were applied to three investigational HIV vaccine trials: HVTN 702 \citep{gray2021vaccine}, HVTN 100 \citep{bekker2018subtype} and HVTN 097 \citep{gray2019immune}. The vaccine regimen used in HVTN 097 is ALVAC-HIV-vCP1521 + subtypes B/E gp120 protein with alum adjuvant, labeled as $\text{P}_{AE/B}/\text{alum}$; the vaccine regimen used in HVTN 100 and HVTN 702 is ALVAC-HIV-vCP2438 + subtype C gp120 protein with MF59 adjuvant, labeled as $\text{P}_{C}/\text{MF59}$. 
Our analysis characterized the immunogenicity of these vaccines in terms of their impact on CD4+ T cells expressing cytokines in response to three antigens: ZM96, TV1 and 1086. The percentage of CD4+ T cells expressing cytokines in response to antigen are measured by the intracellular cytokine staining (ICS) assay. We evaluated the readout of this assay as both a continuous response magnitude and a binary response (0: Yes, 1: No), the latter indicating that the assay readout met the criteria positivity. 
The analysis adjusted for the following baseline participant characteristics: age, sex at birth, BMI, region of enrollment, and educational level. In HVTN 702, the immune responses were measured subject to a case-control sampling scheme with known sampling weights. Participants who were vaccinated in either of the other two trials are assigned weight one since the target immune markers are all measured. 

We present results that compare the difference between vaccine $\text{P}_{AE/B}/\text{alum}$ evaluated in HVTN 097 versus vaccine $\text{P}_{C}/\text{MF59}$ evaluated in HVTN 100 and HVTN 702 among HVTN 702 population where the most prevalent HIV subtype in South Africa is clade C. To provide a benchmark for comparison, we included unadjusted estimators based on the sample average immune response within each trial arm.

Table~\ref{tab:hiv_case} displays results for both estimators of the positive response rate (RR) of CD4+ T cells expressing two cytokines in response to a specifc antigen. Comparing RR between HVTN 097 and HVTN 702 for antigen ZM96, the unadjusted estimated difference in average RR is 0.097 (-0.055  0.249), which is smaller than the results obtained from TMLE, which is 0.182 (-0.014, 0.365). The geometric average response magnitudes show similar trends: 1.169 (0.844, 1.618) for the raw method and 1.470 (0.964, 2.241) using TMLE. For HVTN 702 population, the TMLE analysis revealed a statistically significant decrease in the CD4+ T-cell RR (0.287 (0.085, 0.466), $p=0.006$) and geometric mean (1.935 (1.275, 2.938), $p=0.002$) of the immune response to antigen 1086 for the vaccine $\text{P}_{C}/\text{MF59}$ compared to the vaccine $\text{P}_{AE/B}/\text{alum}$, but our analysis showed no evidence of a difference for antigens ZM96 and TV1 based on either the unadjusted or TMLE. 

\begin{table}
\centering
\caption{\textbf{The difference in the average immune responses of CD4+ cells between vaccine administered in HVTN 702 and HVTN 097 within the referent population HVTN 702.} The point estimate for each vaccine regimen employed in the reference population HVTN 702 was provided in the second and third columns.
Comparisons were summarized for unadjusted approaches and our proposed method for both contrasts. The contrasts pertain to the difference between response rates and the ratio of geometric means. ICS: Intracellular cytokine staining; RR: response rate; GM: geometric mean.} 
\label{tab:hiv_case}
\begin{tabular}{lccc}
  \hline
  Trial & HVTN 702\&100 & HVTN 097 & Difference/Ratio \\
  \hline
  Vaccine & $\text{P}_{C}/\text{MF59}$ & $\text{P}_{AE/B}/\text{alum}$ & -\\
  \hline
  \textbf{Antigen: ZM96} \\
  RR (unadj) & 0.639 & 0.736 & 0.097 \\
  CI & (0.527, 0.752) & (0.634, 0.838) & (-0.055, 0.249) \\  
  p value & -- & -- & 0.210 \\  
  RR (TMLE) & 0.573 & 0.755 & 0.182 \\
  CI & (0.472, 0.668) & (0.559, 0.882) & (-0.014, 0.365) \\  
  p value & -- & -- & 0.068 \\
  GM (unadj) & 0.077 & 0.090 & 1.169 \\
  CI & (0.060, 0.099) & (0.073, 0.112) & (0.844, 1.618) \\  
  p value & -- & -- & 0.348 \\  
  GM (TMLE) & 0.074 & 0.109 & 1.470 \\
  CI & (0.060, 0.092) & (0.076, 0.158) &  (0.964, 2.241)\\  
  p value & -- & -- & 0.074 \\
  \hline
  \textbf{Antigen: TV1} \\
  RR (unadj) & 0.723 & 0.736 & 0.013 \\
  CI & (0.618, 0.827) & (0.634, 0.838) & (-0.133, 0.160) \\  
  p value & -- & -- & 0.857 \\  
  RR (TMLE) & 0.649 & 0.755 & 0.106 \\
  CI & (0.554, 0.734) & (0.559, 0.882) & (-0.083, 0.288) \\  
  p value & -- & -- & 0.272 \\
  GM (unadj) & 0.080 & 0.090 & 1.129 \\
  CI & (0.063, 0.101) & (0.073, 0.112) & (0.819, 1.557) \\  
  p value & -- & -- & 0.458 \\  
  GM (TMLE) & 0.076 & 0.109 & 1.441 \\
  CI & (0.062, 0.093) & (0.076, 0.158) & (0.948, 2.190) \\  
  p value & -- & -- & 0.087 \\
  \hline
  \textbf{Antigen: 1086} \\
  RR (unadj) & 0.546 & 0.736 & 0.190 \\
  CI & (0.430, 0.662) & (0.634, 0.838) & (0.035, 0.344) \\  
  p value & -- & -- & 0.016 \\  
  RR (TMLE) & 0.468 & 0.755 & 0.287 \\
  CI & (0.370, 0.569) & (0.559, 0.882) & (0.085, 0.466) \\  
  p value & -- & -- & 0.006 \\
  GM (unadj) & 0.060 & 0.090 & 1.510 \\
  CI & (0.048, 0.075) & (0.073, 0.112) & (1.103, 2.067) \\  
  p value & -- & -- & 0.010 \\  
  GM (TMLE) & 0.057 & 0.109 & 1.935 \\
  CI & (0.046, 0.069) & (0.076, 0.158) & (1.275, 2.938) \\  
  p value & -- & -- & 0.002 \\
  \hline
\end{tabular}
\end{table}

\section{Discussion}
\label{sec:discussion}
Our framework explicitly outlines sufficient assumptions for a causal interpretation of vaccine immunogenicity comparison across trials. We acknowledge that these assumptions are strong and may not be justifiable in many practical applications, particularly the assumption of ignorability of trial enrollment given measured covariates. This may limit the interpretability of our results in the language of counterfactuals. Nevertheless, we argue that transparently explicating these assumptions is critical. This clarity allows researchers to understand whether and how we should adjust for covariates when calculating average immune responses. Moreover, we argue that the observed data parameter $\psi_X(a)$ is likely to be closer to the true counterfactual immune response $\psi_{\bm{X}}(a)$ than the na{\"i}ve estimand $\mu_a$. Thus, an interpretation of $\psi_X(a)$ as a covariate-standardized immune response may prove satisfactory for advancing scientific aims even in settings where the assumptions required for full counterfactual interpretation fail.

Our framework leads to at least two practical recommendations for the conduct of vaccine immunogenicity studies. First, because the assumptions required for a causal comparison hinges on the availability of key covariates across \emph{all trials} in which the vaccines of interest are evaluated, standardizing the collection of covariates across different trials should be a priority. While fully standardizing a set of covariates across trials with different sponsors or vaccine developers may be unrealistic, funding organizations and vaccine trial networks may consider developing standardized operating procedures for covariate collection for all studies of candidate vaccines. This may be particularly important for pathogens for which there is considerable prior exposure in the trial population. In these instances, pre-existing immunity may significantly modify the immunogenicity of vaccines and may differ across trials. Therefore, every effort should be made to standardize the assays that measure pre-existing immunity in each trial. We discuss other avenues for relaxing assumptions related to covariate availability in Supplementary Section F.
 
A second practical recommendation suggest by our framework is that it may be desirable to publish covariate-conditional estimates of vaccine immunogenicity. While our framework has focused on nonparametric estimation in settings where data from all trials are simultaneously available to the analyst, our results suggest that publishing even simple covariate-conditional immunogenicity models (e.g., based on logistic regression) may lead to more objective immunogenicity comparisons. Moreover, such models may help stimulate new hypotheses pertaining to individual-level factors that influence vaccine immunogenicity. We argue that reporting conditional models may be particularly important for performing improved meta analyses for establishing vaccine correlates of protection. In such analyses, it is typical to visually depict estimated VE from several trials along the vertical axis plotted against some marker of vaccine-induced immunogenicity on the horizontal axis. If the immune response in question is a strong candidate for a correlate of protection, we would expect a positive correlation between vaccine-induced immune responses and protective efficacy of the vaccine. However, we note that these trials themselves may be conducted across diverse populations and thus we should consider standardizing not only the vaccine-induced immunogenicity readout, but also the estimated VE readout from the trial, in order to have the most appropriate means of evaluating an immune response as a potential correlate of protection. We hypothesize applying covariate standardization to meta-analyses in certain contexts could enhance the power for detecting correlates of protection.

\backmatter









\bibliographystyle{biom}
\bibliography{ref}

\label{lastpage}

\end{document}